# Local Environment of Ferromagnetically Ordered Mn in Epitaxial InMnAs

**P.T. Chiu, B.W. Wessels** [a)]
*Department of Materials Science and Engineering and Materials Research Center, Northwestern University, Evanston IL 60208*

**D.J. Keavney, J.W. Freeland**
*Advanced Photon Source, Argonne National Laboratory, Argonne, IL 60439*

---

## Abstract

The magnetic properties of the ferromagnetic semiconductor $In_{0.98}Mn_{0.02}As$ were characterized by x-ray absorption spectroscopy and x-ray magnetic circular dichroism. The Mn exhibits an atomic-like $L_{2,3}$ absorption spectrum that indicates that the 3*d* states are highly localized. In addition, a large dichroism at the Mn $L_{2,3}$ edge was observed from 5-300 K at an applied field of 2T. A calculated spectrum assuming atomic $Mn^{2+}$ yields the best agreement with the experimental InMnAs spectrum. A comparison of the dichroism spectra of MnAs and InMnAs show clear differences suggesting that the ferromagnetism observed in InMnAs is not due to hexagonal MnAs clusters. The temperature dependence of the dichroism indicates the presence of two ferromagnetic species, one with a transition temperature of 30 K and another with a transition temperature in excess of 300 K. The dichroism spectra are consistent with the assignment of the low temperature species to random substitutional Mn and the high temperature species to Mn near-neighbor pairs.

---

a) Electronic mail: b-wessels@northwestern.edu

Ferromagnetic alloys composed of a III-V semiconductor doped with Mn have been extensively studied due to their potential use in various spintronic device applications.[1] The realization of a practical spintronic device hinges on the development of a room temperature ferromagnetic semiconductor. Initial efforts have centered on InMnAs films prepared by low temperature molecular beam epitaxy (MBE).[2] Films with Curie temperatures ($T_c$) of the order of 30-50 K were observed and the magnetism was attributed to random substitutional Mn from extended absorption fine structure (EXAFS) measurements. InMnAs MBE films with $T_c$'s over 300 K were also observed, but were found to contain precipitates. EXAFS patterns were modeled assuming the presence of hexagonal MnAs.[3] Recently, it has been shown that InMnAs thin films grown by metal organic vapor phase epitaxy (MOVPE) exhibit a $T_c$ of 330 K.[4,5] In contrast to MBE InMnAs films, recent EXAFS measurements on these high $T_c$ films indicate the presence of short range order involving Mn substituting on nearest neighbor cation sites.[6] No evidence of hexagonal MnAs was observed in the single phase films by EXAFS. The high temperature ferromagnetism in the material was attributed to Mn on adjacent cation sites.

However, the electronic structure of these alloys has not been determined. X-ray absorption spectroscopy (XAS) utilizes the strong dependence of the absorption edge position on atomic number to determine the electronic structure of each alloy component. For InMnAs, the $L_{2,3}$ absorption of Mn results from the excitation of 2*p* core levels to 3*d* valence states, providing a direct measurement of the 3*d* valence structure of Mn. In contrast to previous studies of core-level photoemission spectra, this yields characteristic line shapes in the XAS indicative of a particular electronic ground state.[7] X-ray magnetic circular dichroism (XMCD) adds an additional level of specificity. The dichroism spectral line shape is specific to the local environment of the ferromagnetically ordered Mn ions. Consequently, XMCD can potentially differentiate between ferromagnetism from substitutional Mn and MnAs precipitates.[8] Useful magnetic parameters such as the spin magnetic moment can also be potentially derived from an XMCD spectrum through the application of an atomic model.[9,10] In this paper, we report a study of the x-ray absorption and dichroism of InMnAs films grown by MOVPE.

An epitaxial $In_{0.98}Mn_{0.02}As$ film with a thickness of 240 nm was grown on a GaAs (001) substrate by atmospheric pressure metal organic vapor phase epitaxy at a temperature of 520?C. The growth process has been described previously.[11] The sample was left uncapped to avoid interference with the XMCD signal from an undoped layer. The InMnAs alloy had a zinc blende structure and was nominally phase pure as determined by x-ray diffraction (XRD). The single phase InMnAs film also exhibited room temperature ferromagnetism as indicated by the observation of hysteresis by magneto-optical Kerr effect measurements.[12]

The XAS and XMCD measurements were performed using the soft x-ray beam line (4-ID-C) of the Advanced Photon Source. The beam line provides >96% left or right circularly polarized light using a circular polarized undulator.[13] The sample temperature was controlled in a liquid helium cryostat. The x-ray absorption and dichroism spectra near the Mn $L_{2,3}$ edges were obtained by measuring the total electron yield (TEY) and fluorescence (FY) with the beam incident at 45? from the surface normal. A 2T magnetic field was applied along the beam direction to saturate the sample magnetization. Dichroism scans were taken by switching the photon helicity at each energy point.

Fig. 1a shows the *L*-edge XAS measured by TEY for Mn in an $In_{0.98}Mn_{0.02}As$ thin film taken at 50 K. The two groups of peaks shown in the Mn absorption spectrum correspond to core level transitions from the $2p_{3/2}$ ($L_3$ edge) and $2p_{1/2}$ ($L_2$ edge) states to the 3*d* state. In particular, the Mn $L_3$ absorption is comprised of peaks A, B, and C, and the $L_2$ absorption is composed of peaks D and E. A similar line shape was detected by FY, which suggests that the XAS at the surface is comparable to that of the bulk. The rich structures observed in the spectrum are indicative of a highly localized $Mn^{2+}$ ion with a $3d^5$ ground state configuration.[14] In order to determine the contributions of Mn ions with ground state configurations other than $3d^5$, the absorption spectrum was first normalized to a constant step like background, resulting from transitions into the continuum. The resulting spectrum was subsequently compared with a simulated spectrum based upon a linear combination of atomic Mn spectra with ground state configurations of $3d^4$, $3d^5$, and $3d^6$.[15] The InAs crystal field strength was included as a parameter in the model used. As shown in Fig. 2a, the observed peak positions and line shape in the experimental spectrum are best reproduced for a calculated spectrum using a pure $3d^5$ ground state configuration with a weak crystal field strength of 10 $Dq$=0.5 eV. The quality of the fit confirms that the Mn *d*-electrons have $3d^5$ ground state configuration and are highly localized. In addition, the lack of hybridization of the Mn 3*d* orbitals observed in the XAS indicates that *p*-*d* and *s*-*d* exchange effects are limited. Further investigations of the induced magnetic moments on In and As are needed to clarify the nature of the exchange interactions in MOVPE InMnAs.

Since Mn ions are in the $3d^5$ state, they are essentially fully ionized. In the case that Mn acts as a shallow acceptor, the expected hole concentration is $4 \times 10^{20}$ $cm^{-3}$. This is in contrast to the measured hole concentration of $1 \times 10^{18}$ $cm^{-3}$. The large disparity in hole concentrations suggests that the holes are predominantly bound to the Mn pairs and do not contribute to sample conductivity. The electronic configuration is of the form of $3d^5+h$. In support of this proposal, ab initio calculations have confirmed that Mn dimers form a bonding-antibonding pair that results in increased hole localization.[17] In contrast, randomly substituted Mn ions are shallow acceptors and are the source of the itinerant holes that are responsible for the film conductivity.

The dichroism spectra for the InMnAs film at 5 K and 50 K are shown in the bottom panel of Fig 1b. The dichroism spectra at the two temperatures have the same rich structure, exhibiting features closely related to the absorption spectrum. The peak dichroism occurs at 638.7 eV with a magnitude of 15%, as compared to the theoretical magnitude of 58%.[10] The low dichroism asymmetry is tentatively attributed to a surface oxide formed at the InMnAs film surface.[18] Next, the experimental dichroism of InMnAs at both 5 and 50 K was fitted with respect to a theoretical XMCD spectrum, generated assuming linear combinations of atomic Mn states. As shown in Fig. 2b, the best fit was obtained using the same ground state configuration of $3d^5$ and a crystal field of 10 $Dq$=0.5 eV found in the XAS fitting. The fit indicates that the ferromagnetically ordered Mn are in the high spin $Mn^{2+}$ state and highly atomic at both 5 and 50 K.

The main difference between the dichroism at 5 K as compared to that at 50 K is the magnitude of features A–E. The full dependence of the dichroism magnitude at 638.7 eV as a function of temperature is shown in Fig. 3. In Fig. 3, there is a slow linear decrease in the XMCD with respect to temperature from 50 K to 300 K. For temperatures below 30 K, there is a sharp increase, with the dichroism nearly doubling over a range of 25 K. The monotonic dependence of the XMCD signal vs. $T$ from 30–300 K is consistent with dichroism due to a single ferromagnetic species. However, the abrupt increase at 30 K presumably is the result of a second ferromagnetic species adding to the dichroism.

To determine whether either of the ferromagnetic species are hexagonal MnAs precipitates, the dichroism spectra of the InMnAs taken at 5 K and 50 K were compared to the dichroism of a 100 nm thick MnAs films grown on Si (001). The XAS of both MnAs and InMnAs samples were taken consecutively in order to calibrate the energy scales. As evident in Fig. 4, there are three main points of contrast between the dichroism of MnAs and InMnAs. First, feature A in the XMCD is a sharp negative peak in the case of MnAs while only a subtle negative shoulder for InMnAs. Second, the positive peak corresponding to feature B for InMnAs is absent for MnAs. Finally, the $L_2$ dichroism peaks of MnAs are broadened and red shifted by 1 eV relative to the InMnAs $L_2$ peaks (features C and D). Similar points of contrast are evident in a comparison of the dichroism of InMnAs at 50 K with the MnAs spectrum. Consequently, the significant differences in the $L_3$ dichroism line shape strongly suggest that the low or high temperature ferromagnetic species in the InMnAs film cannot be attributed to hexagonal MnAs precipitates.[8]

Instead, the additional dichroism at low temperatures is attributed to a ferromagnetic species consisting of simple substitutional Mn acceptors. This is consistent with $T_c$'s of less than 50 K observed in MBE grown InMnAs films, where the magnetism was ascribed to randomly substituted Mn.[2] The observation of a low temperature ferromagnetic species in the XMCD also agrees with the observation of negative magneto-resistance only at temperatures less than 14 K.[19] The origin of the negative magnetoresistance is also attributed to random substitutional Mn acceptors. Finally, the atomic $Mn^{2+}$ dichroism line shape observed for InMnAs, has been previously attributed to singly substitutional Mn.[8,20]

The similarity in line shape of the XMCD at 5 and 50 K is consistent with the assignment of one ferromagnetic species to singly substituted Mn and one ferromagnetic species to the Mn dimers (or trimers) previously observed by EXAFS.[6] Randomly substituted Mn ions and Mn dimers are both tetrahedrally coordinated to four As atoms in the nearest neighbor shell. Differences in the local chemical environment of Mn for these two species occur in the second nearest neighbor shell. On the other hand, Mn in hexagonal MnAs is coordinated with six As nearest neighbors. The difference between Mn in InMnAs and Mn in hexagonal MnAs occurs in the *nearest* neighbor shell. Since the differences in chemical environment of the first nearest neighbor shell have a greater effect on the XMCD than differences in the second, contrasting XMCD spectra between InMnAs and hexagonal MnAs should be observed. A comparison of the XMCD of MnAs and InMnAs supports this prediction. Due to the similarities in the first nearest neighbor shells, the two InMnAs ferromagnetic species have matching XMCD spectra.

In summary, we find from XAS that the Mn *d*-electrons in the ferromagnetic semiconductor InMnAs are highly localized and have a $3d^5$ ground state configuration. Strong dichroism is observed from 5 K to 300 K. The temperature dependence of the XMCD indicates the presence of two ferromagnetic species. One species contributes to the dichroism from 30-300 K, while a second low temperature species adds to the dichroism from 5-30 K. A comparison of the dichroism spectra of InMnAs and MnAs at several temperatures suggests that neither ferromagnetic species is hexagonal MnAs. Rather, the dichroism data is consistent with the previous attribution of the high temperature ferromagnetic species to Mn dimers (or trimers), and the low temperature species to random substitutional Mn. A comparison between the experimental and theoretical dichroism demonstrates both species are composed of highly localized $Mn^{2+}$ in a high spin state configuration.

The authors would like to thank J. H. Song of Northwestern University for providing the MnAs/Si sample and Prof. M. van Veenendaal of Northern Illinois University for providing the atomic multiplet code and for helpful discussions. This work is supported by the NSF under the Spin Electronics Program #ECS-0224210 and the MRSEC program under contract DMR-0076797. Use of the Advanced Photon Source was supported by the U.S. Department of Energy, Office of Science, under Contract No. W-31-109-Eng-38.

**Figure Captions:**

Fig. 1: (a) Absorption spectrum of $In_{0.98}Mn_{0.02}As$ measured by TEY at 50 K. The absorption spectrum was normalized relative to the $L_3$ resonance at 638.7 eV. (b) Dichroism at 5 and 50 K at an applied field of 2 T.

Fig. 2: Comparison of experimental XAS and XMCD with theoretical spectra calculated assuming atomic Mn states. (a) Fit of the absorption spectrum. (b) Fit of the dichroism spectra.

Fig. 3: XMCD at 638.7 eV versus temperature of InMnAs with an applied field of 2 T.

Fig. 4: Comparison of the XMCD of $In_{0.98}Mn_{0.02}As$ and MnAs. The dichroism spectrum of InMnAs was taken at 5 K with an applied field of 2 T. The dichroism spectrum MnAs was taken at 300 K.

## Figures

Fig. 1

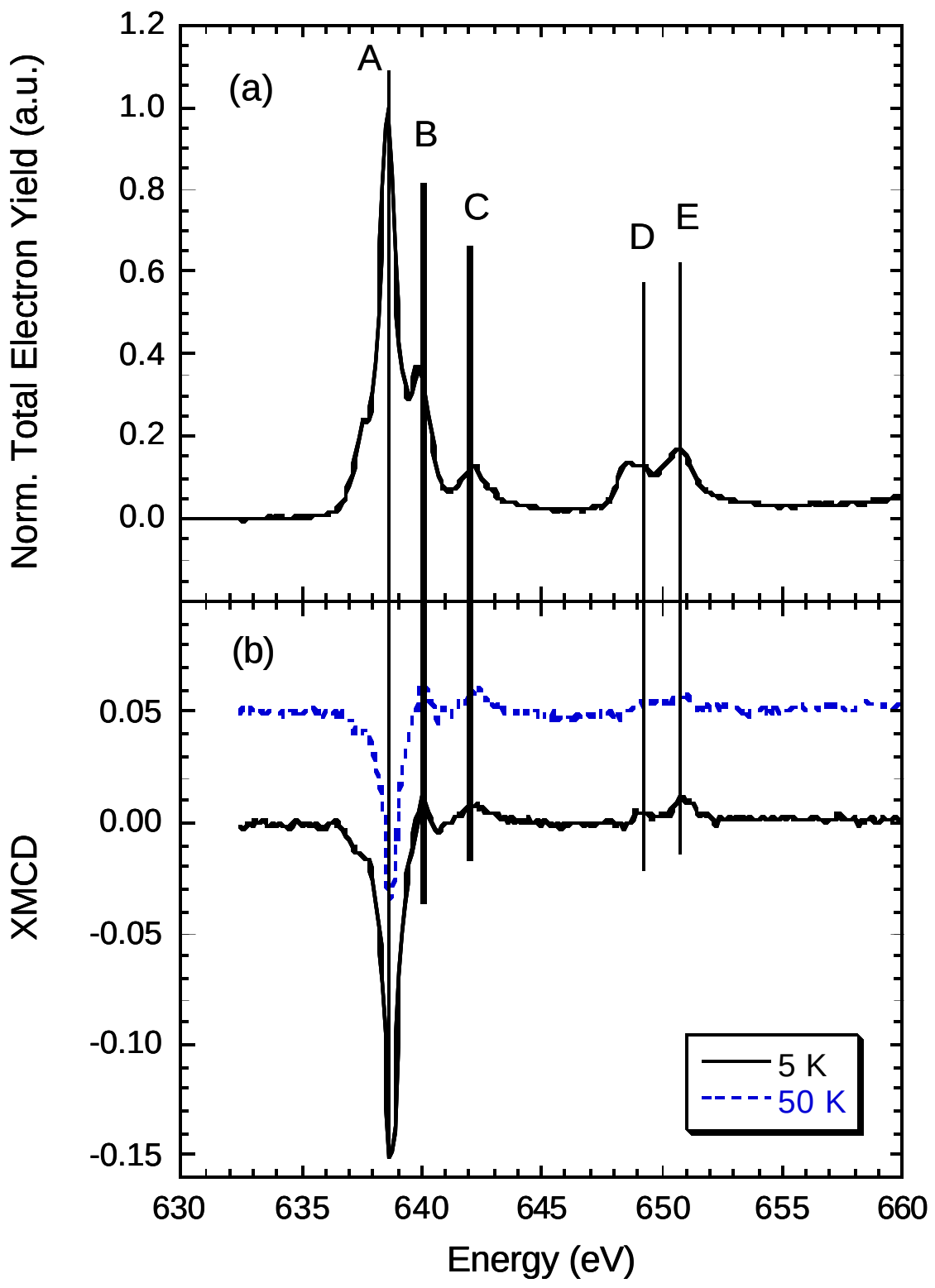

Fig. 2

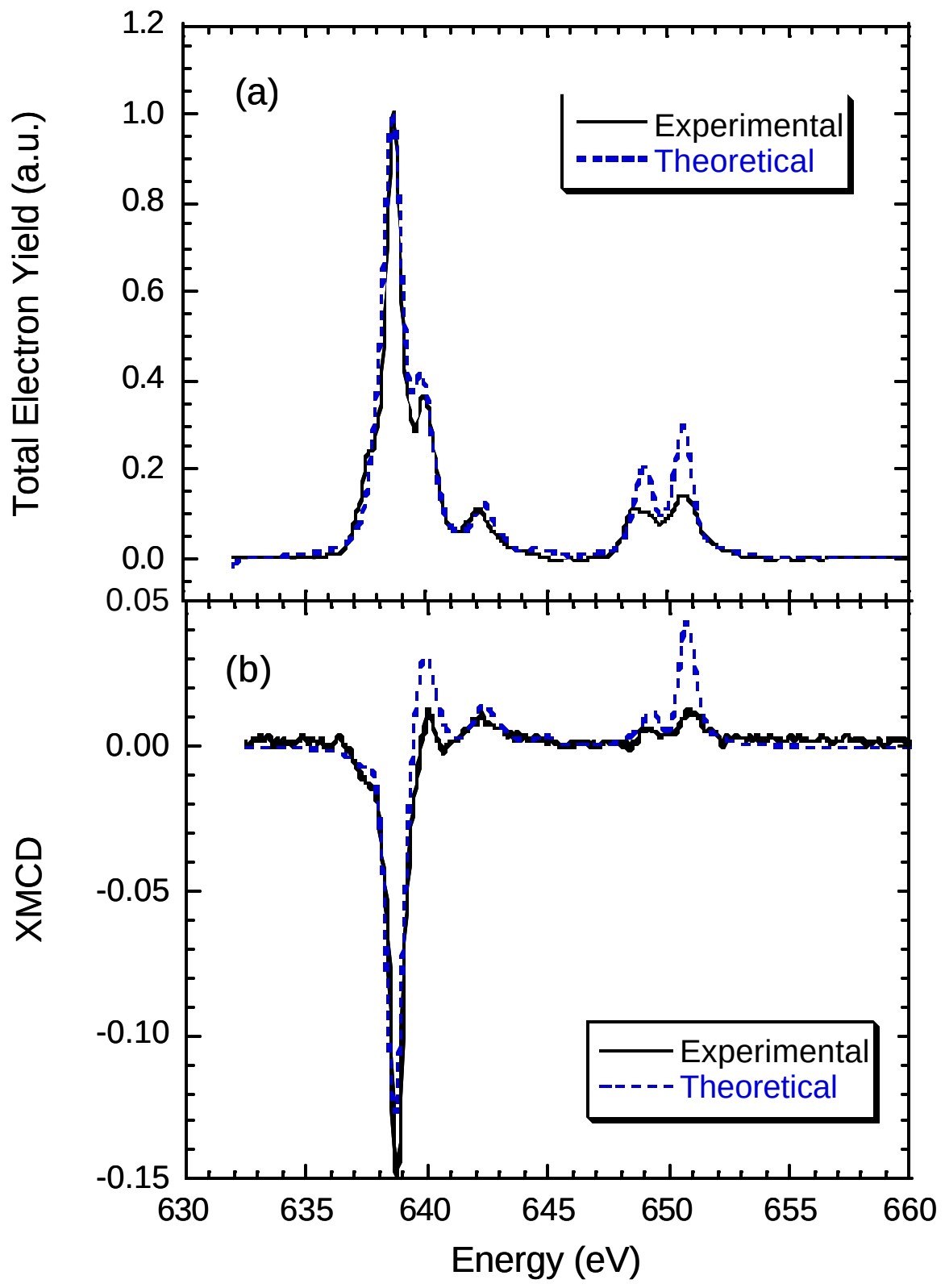

Fig. 3

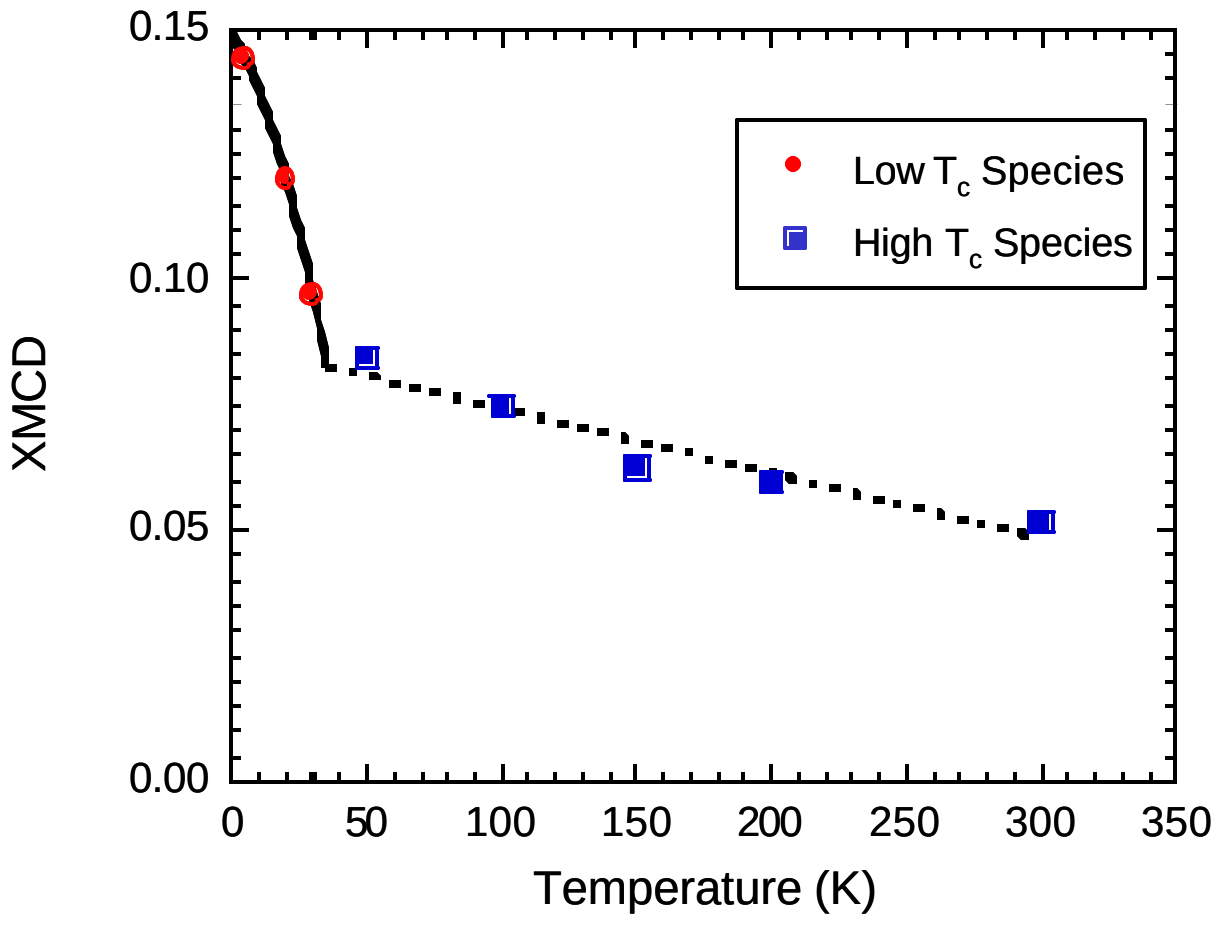

Fig. 4

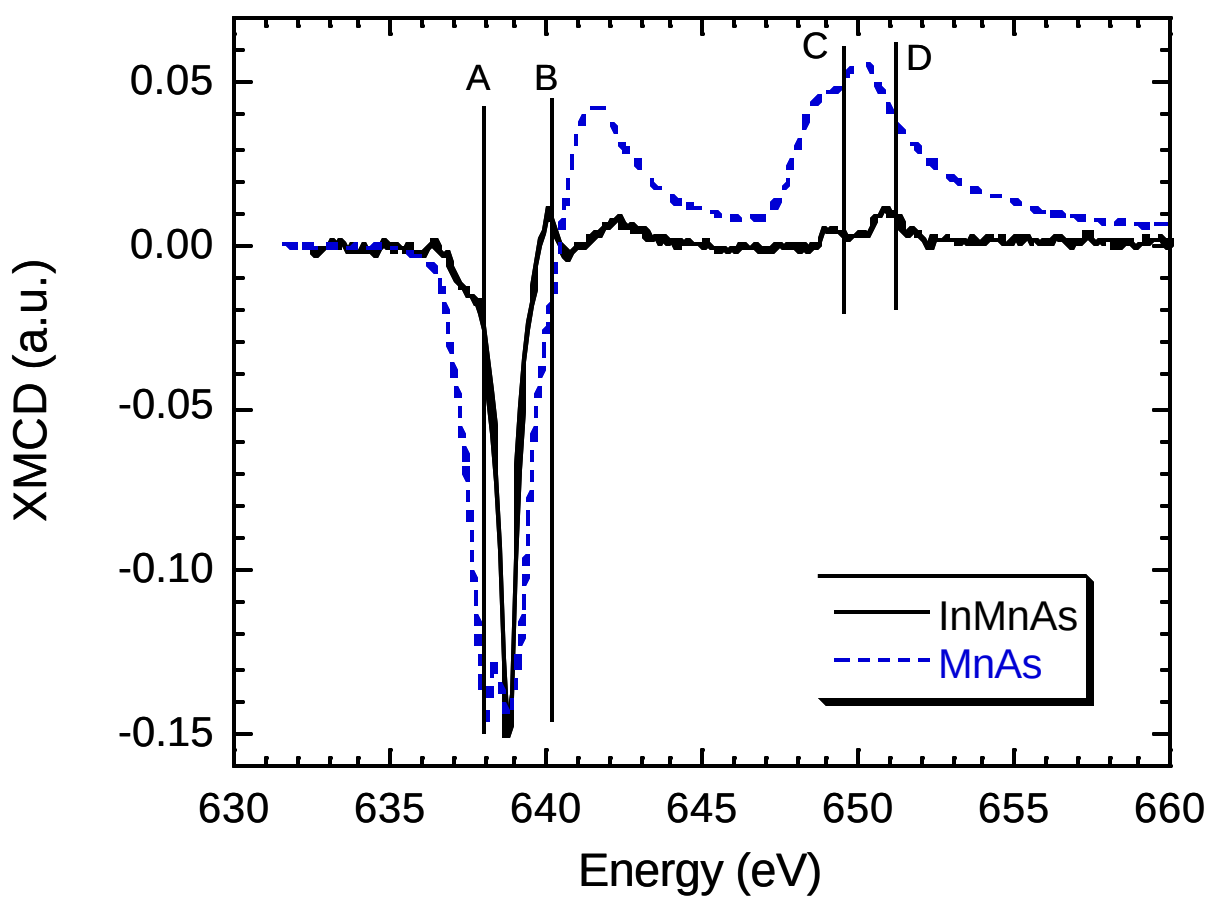